\documentclass[showpacs,amsmath,
twocolumn,
aps,prl]{revtex4}
\usepackage{graphicx}
\usepackage{dcolumn}

\begin{document}

\title{Antiferromagnetic $s-d$ exchange coupling in GaMnAs}
\author{R. C. Myers}
\author{M. Poggio}
\author{N. P. Stern}
\author{A. C. Gossard}
\author{D. D. Awschalom}
\affiliation{Center for Spintronics and Quantum Computation, 
University of California, Santa Barbara, CA 93106}
\date{\today}

\begin{abstract}
  Measurements of coherent electron spin dynamics in
  Ga$_{1-x}$Mn$_{x}$As/Al$_{0.4}$Ga$_{0.6}$As quantum wells with $
  0.0006\% < x < 0.03\% $ show an antiferromagnetic (negative)
  exchange bewteen s-like conduction band electrons and electrons
  localized in the d-shell of the Mn$^{2+}$ impurities. The magnitude
  of the $s-d$ exchange parameter, $ N_{0} \alpha$, varies as a
  function of well width indicative of a large and negative
  contribution due to kinetic exchange. In the limit of no quantum
  confinement, $ N_{0} \alpha$ extrapolates to $-0.09\pm0.03$ eV
  indicating that antiferromagnetic $s-d$ exchange is a bulk property
  of GaMnAs.  Measurements of the polarization-resolved
  photoluminescence show strong discrepancy from a simple model of the
  exchange enhanced Zeeman splitting, indicative of additional
  complexity in the exchange split valence band.
\end{abstract}

\maketitle

The $sp-d$ exchange constants in II-VI dilute magnetic semiconductors
(DMS) are readily measurable through magneto-optical spectroscopy;
however, the high concentration of growth defects in
Ga$_{1-x}$Mn$_{x}$As ($x > 1\%$) has precluded band-edge optical
measurements of its exchange splitting.  Despite this constraint,
several estimates of the $p-d$ exchange constant ($N_{0} \beta$) have
been published from modeling of transport \cite{Matsukura:1998},
core-level photoemission \cite{Okabayashia:1998}, and cyclotron
resonance measurements \cite{Zudov:2002}.  A previous study on highly
dilute Ga$_{1-x}$Mn$_{x}$As ($x < 0.1\%$) crystals grown by the
Czochralski method reports polarized magnetoreflectance data from
which the total exciton spin splitting is determined to within 600 meV
\cite{Szczytko:1996}.  This result, however, includes no independent
measurements of the $s-d$ exchange constant ($N_{0} \alpha$) or
$N_{0}\beta$; the reported estimation of $N_{0} \beta$ depends on an
assumed positive value of $N_{0} \alpha$ based on work in II-VI DMS.

Here we investigate MBE grown
Ga$_{1-x}$Mn$_{x}$As/Al$_{0.4}$Ga$_{0.6}$As quantum wells (QWs) with $
0.0006\% < x < 0.03\% $ in which Mn$^{2+}$ ions act as spin 5/2
paramagnets. Ga$_{1-x}$Mn$_{x}$As is typically grown at temperatures
around 250$^\circ$C, however, for these values of $x$, the growth
temperature can be increased to 400$^\circ$C while still allowing
substitutional incorporation of Mn. The use of this increased
substrate temperature is seen to reduce excess As in our samples and
enables the observation of magneto-optical effects such as
polarization resolved photoluminescence (PL) and Kerr rotation (KR)
\cite{Poggio:tbp}. Both the sign and magnitude of $N_{0}\alpha$ are
measured over a wide range of QW widths and Mn concentrations by
time-resolved measurements of KR.  In contrast to II-VI DMS, we find
that $N_{0}\alpha < 0$ in all QWs measured. A measured negative trend
in $N_0 \alpha$ as a function of decreasing QW width is consistent
with the mechanism of kinetic exchange for conduction band electrons
observed in II-VI DMS QWs \cite{Merkulov:1999}. Using a fit to this
model in the limit of wide wells, we extrapolate a negative $N_0
\alpha$ for bulk GaMnAs. The change in $N_0 \alpha$ due to quantum
confinement is as large as $-740$ meV, much larger than observed by
Merkulov \textit{et al.} in II-VI DMS. Measurements of
polarization-resolved PL reveal strong non-linearities in the Zeeman
splitting, making it difficult to extract the exchange-induced exciton
energy $N_{0}(\alpha - \beta)$, and thus $N_{0}\beta$.

The samples consist of single Ga$_{1-x}$Mn$_{x}$As QWs of width $d$
surrounded by Al$_{0.4}$Ga$_{0.6}$As barriers and are grown on (001)
semi-insulating GaAs wafers.  A detailed discussion of sample growth
and quantitative measurements of $x$ by secondary ion mass
spectroscopy are provided elsewhere \cite{Poggio:tbp}.

Electron spin dynamics are measured by time-resolved KR with the
optical axis perpendicular to the applied magnetic field $B$ (Voigt
geometry) and parallel to the growth direction $\hat{z}$. The
measurement, which monitors small rotations in the linear polarization
of laser light reflected off of the sample, is sensitive to the spin
polarization of electrons in the conduction band of the QW
\cite{Crooker:1997}.  A Ti:Sapphire laser with a 76-MHz repetition
rate and 250-fs pulse width tuned to a laser energy $E_L$ near the QW
absorbtion energy is split into a pump (probe) beam with an average
power of 2 mW (0.1 mW).  The helicity of the pump beam polarization is
modulated at 40 kHz by a photo-elastic modulator, while the intensity
of the linearly polarized probe beam is modulated by an optical
chopper at 1 kHz for lock-in detection.  Both beams are focused to an
overlapping 50 $\mu$m diameter spot on the sample which is mounted
within a magneto-optical cryostat. The time delay $\Delta$t between
pump and probe pulses is controlled using a mechanical delay line. The
pump injects electron spins polarized perpendicular to $B$ into the
conduction band of the QW. The change in the probe polarization angle,
$\theta _K \left( {\Delta t} \right)$ is proportional to the average
electron spin polarization in the QW and is well fit to a single
decaying cosine, $\theta _K \left( {\Delta t} \right) = \theta_{\perp}
e^{ - \Delta t / T_2 ^\ast }\cos \left( {2\pi \nu _L \Delta t + \phi }
\right)$, where $\theta_{\perp}$ is proportional to the total spin
injected, $T_2 ^\ast$ is the inhomogeneous transverse spin lifetime,
$\nu _L$ is the electron spin precession (Larmor) frequency, and
$\phi$ is the phase offset. No evidence of Mn$^{2+}$ spin precession,
which occurs in II-VI magnetically doped QWs \cite{Crooker:1997}, has
been observed in the samples studied here.  The order of magnitude
smaller $x$ in our III-V QWs compared to the II-VI QWs puts any
Mn$^{2+}$ spin precession signal below the experimental detection
limit.

Fig. 1a shows typical time-resolved KR data measured at $B = 8$ T for
a Mn-doped QW ($d=7.5$ nm and $x\sim0.01\%$) together with fit, as
described above, demonstrating electron spin coherence in the GaMnAs
system.  $\nu _L$ is proportional to the total conduction band spin
splitting between spin-up and spin-down electrons ($\Delta
E=E\uparrow-E\downarrow$) and can be expressed in terms of the Zeeman
splitting ($\Delta E _g$), and the $s-d$ exchange splitting ($\Delta
E_{s - d}$):
\begin{equation}
\label{eq1}
h \nu_L = \Delta E = \Delta E _g + \Delta E_{s-d} = 
g _e \mu _B B - x N _0 \alpha \langle S_x \rangle.
\end{equation}
Here $h$ is Planck's constant, $g _e$ is the in-plane electron
g-factor, $\mu _ B$ is the Bohr magneton, and $\langle S_x \rangle$ is
the component of Mn$^{2+}$ spin along $B$.  We emphasize that a
measurement of $\nu_L$ alone, because of phase ambiguity, does not
determine the sign of $\Delta E$.  $\langle S_x \rangle = -
\frac{5}{2} B_{5/2} \left ( \frac{5 g_{Mn} \mu_B B}{2 k_B
    (T-\theta_P)} \right )$, where $B_{5/2}$ is the spin-5/2 Brillouin
function, $g_{Mn}$ is the g-factor for Mn$^{2+}$, $k_{B}$ is
Boltzman's constant, and $\theta_P$ is the paramagnetic Curie
temperature. Note that since the g-factor for Mn$^{2+}$ ($g_{Mn}=2$)
is positive, then for $B>0$, $\langle S_x \rangle < 0$.

In Fig. 1b, $\nu_L$ is plotted as a function of $B$ for a set of four
samples with $d=7.5$ nm and varying $x$.  The non-magnetic ($x=0$)
sample shows a linear field dependence of $\nu_L$, from which we
extract values of $g_e$ as described in Eq. (\ref{eq1}). As the Mn
doping concentration is increased, $\nu_L$ increases and its $B$
dependence becomes non-linear. Further, this field dependence shows
the same Brillouin function behavior that is expected for the
magnetization of paramagnetic GaMnAs, Eq. (\ref{eq1}).
\begin{figure}[b]\includegraphics{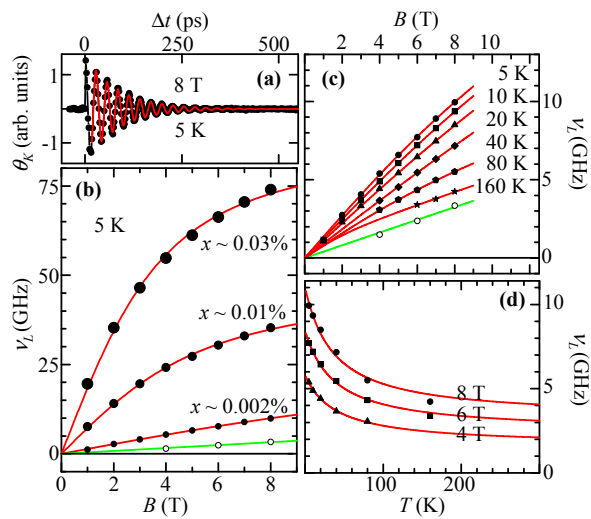}\caption{\label{fig1}
    (Color) Time-resolved electron spin dynamics in $d=7.5$ nm GaMnAs
    QWs. (a) An example of KR data (points) together with fit (line).
    (b) $\nu_L$ as a function of $B$ for different $x$ values (solid
    points); larger points indicate increasing $x$. Open data points
    are for the $x=0$ sample. (c) The effect of increasing $T$ on the
    $B$ dependence of $\nu_L$ for the sample with $x\sim0.002\%$
    (solid points) and for the $x=0$ sample (open points).  (d) $T$
    dependence of $\nu_L$ at constant $B$ for the $x\sim0.002\%$
    sample. Red lines in (b-d) are fits to Eq. (\ref{eq2}).
  }\end{figure}

The dependence of $\nu_L$ on $B$ and $T$ for the sample with $d = 7.5$
nm and $x\sim0.002\%$ is plotted in Fig. 1c and d together with values
for the control sample, $x=0$ and $d=7.5$ nm. For the magnetic sample,
as $T$ is increased $\nu_L$ decreases asymptotically toward the
control sample value $g_e \mu_B B / h$ without crossing zero (Fig.
1c). Thus, it follows from Eq. (\ref{eq1}) and from the sign of
$\langle S_x \rangle$ that for $d = 7.5$ nm $N_0 \alpha$ has the same
sign as $g_e$.  For $d = 7.5$ nm, $g_e < 0$ \cite{Snelling:1991}, and
thus $N_0 \alpha < 0$. This conclusion is also supported by the QW
width dependence discussed below.

Using $g_e$ extracted from the $x = 0$ sample (green line) and
Eq. (\ref{eq1}) we fit $\nu_L$ as a function of $B$ and $T$ to,
\begin{equation}
\label{eq2}
\nu_L = \frac{g _e \mu _B B}{h} +\frac{5 A}{2 h} B_{5/2} \left ( \frac{5 \mu_B B}{k_B (T - \theta_P)} \right ),
\end{equation}
which has only two fit parameters, $A$ and $\theta_P$. Comparing Eqs.
(\ref{eq2}) to (\ref{eq1}), it is clear that $ A = x N_0 \alpha $. The
data in Fig. \ref{fig1}b-d are fit to Eq. (\ref{eq2}), with fits shown
as red lines.  A large negative $\theta_P$ (-37 K) is extracted from
the fits for the sample with the lowest Mn doping (Fig.1c and d),
which may be explained by an increased spin temperature of Mn$^{2+}$
due to photoexcitation. This effect has been reported in II-VI DMS for
low magnetic doping levels \cite{Keller:2001}. Also supporting this
hypothesis, we find smaller values of $|\theta_P|<7$ K in samples with
larger $x$.

$N_0 \alpha$ is examined in detail for QWs of varying $d$.  For this
analysis, four sets of samples with various $x$ (including $x = 0$)
were grown for $d = 3, 5, 7.5$ and $10$ nm. Note that each sample set
of constant $d$ was grown on the same day, which we have observed to
reduce QW thickness variations between samples within each set from
$\sim3$\% to $<1$\%.  In Fig. 2a, $g_e$ in the non-magnetic ($x=0$)
QWs is plotted as a function of $d$ together with data from two other
publications \cite{Snelling:1991,Poggio:2004}. Our data track the
thickness dependence of the QW g-factor as previously reported with a
slight positive shift in $g_e$.  The larger Al concentration (40\%) in
the QW barriers used in our samples versus the concentration (33\%)
used in Refs.  \cite{Snelling:1991,Poggio:2004} accounts for this
discrepancy \cite{Weisbuch:1977}.  Knowing the absolute sign of $g_e$
for QWs of any width, we determine the sign of $N_0 \alpha$ for each
$d$ in the manner described previously. With a now calibrated sign,
$\Delta E = h \nu_L$ is plotted in Fig.  2b as a function of $B$ for
all four QW sample sets with varying $d$.  As shown in Fig. 2b, for
any given $d$, $\Delta E$ decreases as $x$ increases. Following from
Eq. (\ref{eq1}) and from the sign of $\langle S_x \rangle$, this
demonstrates that $N_0 \alpha$ is negative, i.e.  antiferromagnetic,
which has been reproduced unambiguously in at least 20 additional
samples. The effect of increasing temperature on the $B$ dependence of
$\Delta E$ for the $d=5$ nm and $x\sim0.008\%$ sample is shown in Fig.
2c, which dramatically illustrates the negative $s-d$ constant. For $
d = 5$ nm, $g_e$ is weakly positive, thus for $B>0$ and at high
temperature $\Delta E > 0$. As the temperature descreases, $\Delta
E_{s-d}$ becomes more negative as the paramagnetic susceptibility
increases. At $T = 10$ K and $B = 7$ T, $\Delta E = 0$ since the $s-d$
exchange splitting is equal and opposite to the Zeeman splitting. For
lower temperature, $\Delta E < 0$ since $\Delta E_{s-d} > \Delta E
_g$. We note that the data are well fit to Eq. (\ref{eq2}) despite
their highly non-linear nature. We contrast our observation of
antiferromagnetic $s-d$ exchange in III-V GaMnAs, with the
ferromagnetic $s-d$ exchange ubiquitous in II-VI DMS. Their symmetry
forbids hybridization of s and d orbitals, such that only direct
(ferromagnetic) $s-d$ exchange is possible \cite{Larson:1988}.  The
antiferromagnetic $s-d$ exchange in GaMnAs may be due to the narrower
band gap of this material compared with II-VI, such that the
conduction band has partial p-character thus allowing hybridization
with the d orbitals localized on the Mn$^{2+}$ impurities.

\begin{figure}[b]\includegraphics{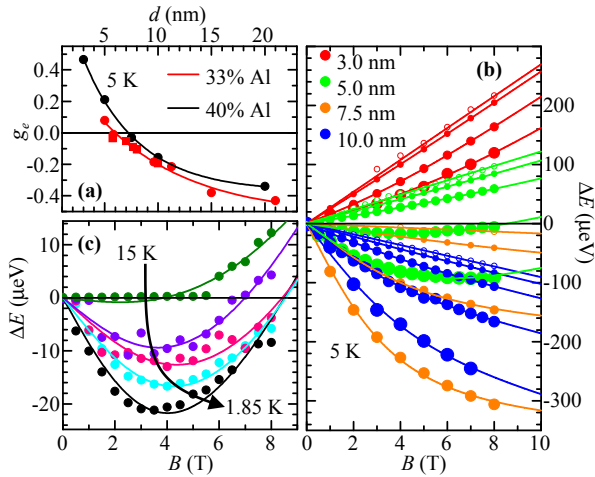}\caption{\label{fig2}
    (Color) (a) $g_e$ as a function of $d$. Black points are from
    control ($x=0$) samples of this study, red circles and squares are
    from references \cite{Snelling:1991,Poggio:2004}, respectively.
    Lines guide the eye. (b) $\Delta E$ as a function of $B$ for QWs
    with (solid circles) and without Mn doping (open circles); larger
    points indicate increasing $x$. (c) $\Delta E$ for the sample with
    $d=5$ nm and $x\sim0.008\%$ at various temperatures.  Fits to Eq.
    (\ref{eq2}) are shown as lines in (b) and (c).  }\end{figure}

In Fig. 3a, the fit parameter $A = x N_0 \alpha$ is plotted as a
function of $x$ together with linear fits for each sample set of
constant $d$. The finite values of $x N_0 \alpha$ at $x = 0$,
extrapolated from the linear fits, are attributed to either the
experimental error in the determination of $g_e$ in the non-magnetic
QWs or error in the measurement of $x$, both errors which to first
approximation have negligible effect on the slope.  These linear fits
demonstrate that $N_0 \alpha$ is constant over the measured doping
range for QWs with the same width, but it varies with $d$ as plotted
in Fig. 3b. $N_0 \alpha$ is more negative the narrower the QW, while
it appears to saturate for wide QWs.  In II-VI DMS QWs, a negative
change in $N_0 \alpha$ as large as $-170$ meV was previously reported
for increasing confinement and was attributed to a kinetic exchange
coupling due to the admixture of valence and conduction band wave
functions \cite{Merkulov:1999}.  We plot $N_0 \alpha$ as a function of
the electron kinetic energy ($E_{e}$) in Fig. 3c, and the data are
linear. Extrapolating to $E_{e}=0$ we obtain a bulk value of $N_0
\alpha=-0.09\pm0.03$ eV for GaMnAs. A change in $N_0 \alpha$ as large
as $-740$ meV is observed in the narrrowest wells measured ($d=3$ nm)
and the slope of $N_0 \alpha(E_{e})$ is $\sim5$ times larger than in
Merkulov \textit{et al}.
\begin{figure}[b]\includegraphics{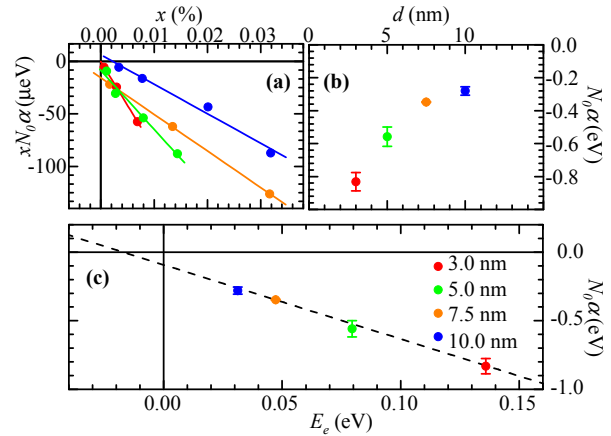}\caption{\label{fig3}
    (Color) (a) $x N_0 \alpha$ as a function of $x$ from fits shown in
    Fig. 2b; error bars are the size of the points. Linear fits are
    shown for each sample set of constant $d$. (b) $N_0 \alpha$
    extracted from fits in (a) and plotted as a function of $d$. (c)
    The electron exchange constant $N_0 \alpha$ as a function of
    electron kinetic energy for GaMnAs.  }\end{figure}

\begin{figure}[b]\includegraphics{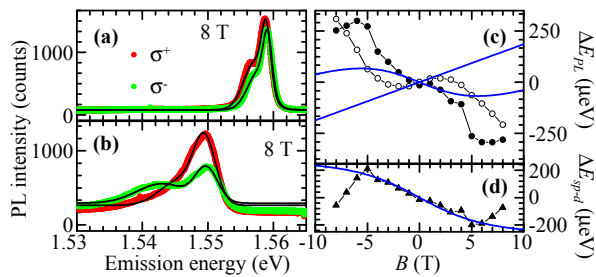}\caption{\label{fig4}
    (Color) Polarization-resolved PL for 10 nm QW at 5 K. (a)
    $\sigma^{+}$ and $\sigma^{-}$ polarized PL from the $x=0$ QW
    excited with 1 W/cm$^2$ at 1.722 eV. 2-Gaussian fits to the data
    are shown as black lines and the higher energy Gausssian is
    attributed to the heavy hole exciton in the QW. (b) Similar data
    shown for the $x\sim0.003\%$ sample excited with 10 W/cm$^2$ at
    1.722 eV.  (c) $B$ dependence of $\Delta E_{PL}$ for the $x=0$ and
    $x\sim0.003\%$ samples plotted as open and filled circles,
    respectively. Fits to $\Delta E_{PL}$ for $|B| < 2$ T are shown
    for both samples as blue lines. (d) The difference between $\Delta
    E_{PL}$ in the $x\sim0.003\%$ sample and in the $x=0$ sample,
    $\Delta E_{sp-d}$, (black triangles) plotted with fit for $|B| <
    5$ T (blue line).  }\end{figure}

Polarization-resolved PL is measured as a function of $B$ in the
Faraday geometry with PL collected normal to the sample surface. The
excitation laser is linearly polarized and focused to a spot 100
$\mu$m in diameter with an energy set above the QW absorption energy.
While PL is seen to quench with increasing Mn doping, QWs without or
with low Mn doping emit PL whose energy dependence is well fit by two
Gaussians (Fig. \ref{fig4}a and b). In the non-magnetic samples, the
emission energy of the narrower, higher-energy Gaussian peak tracks
the $B$ dependence expected for the Zeeman splitting in QWs,
indicating that this peak is due to heavy hole exciton recombination.
On the other hand, the wider, lower-energy Gaussian is likely due to
donor-bound exciton emission from shallow donors in the QWs. These
shallow donors are likely Mn interstitials, since the emission
linewidth increases as the calculated Mn interstitial concentration
increases.  The lower energy Gaussian is also present in the
non-magnetic samples (Fig.  \ref{fig4}a) perhaps due to an impurity
level of Mn interstitials ($\leq 10^{15}$ cm$^{-3}$)
\cite{Poggio:tbp}.

The splitting in the polarized emission energy of the higher energy
Gaussian, $\Delta E_{PL} = E_{\sigma^{+}} - E_{\sigma^{-}}$, is
measured in all the non-magnetic samples. For small fields ($B < 2$
T), $\Delta E_{PL}$ depends linearly on field with the slope giving
the out-of-plane heavy hole exciton g-factor ($g_{ex}$). The extracted
values of $g_{ex}$ agree within the experimental error with previously
published values \cite{Snelling:1992}. At higher fields, $\Delta
E_{PL}$ deviates from linearity, particularly in the wider QWs as
shown in Fig.  \ref{fig4}c where it reverses sign in the 10 nm QW with
$x = 0$ at $|B| = 5$ T.

In Mn-doped samples, $\Delta E_{PL}$ results from both the Zeeman
splitting ($\Delta E_{gex}$) and the $sp-d$ exchange splitting
($\Delta E_{sp-d}$):
\begin{equation}
\label{eq3}
\Delta E_{PL} = \Delta E_{gex} + \Delta E_{sp-d} = - g_{ex} \mu_B B + x N_0 ( \alpha - \beta ) \langle S_x \rangle.
\end{equation}
Using the measurements of $g_{ex}$ from the $x=0$ samples and the
previously extracted values of $\langle S_x \rangle$ and $N_0\alpha$
at 5K (Fig. 2b), we fit $\Delta E_{PL}$ to Eq. (\ref{eq3}). In the 10
nm QW for low fields we estimate $N_0 \beta=-3.4\pm1.5$ eV using the
fits shown in Fig. 4c as blue lines. As Fig. \ref{fig4}c makes clear,
this model breaks down at high fields where non-linearities dominate
$\Delta E_{PL}$. In an attempt to avoid non-linearities in the Zeeman
splitting not modeled by a linear $\Delta E_{gex}$, $\Delta E_{PL}$
measured in the 10 nm $x=0$ sample is subtracted from $\Delta E_{PL}$
of the 10 nm $x\sim0.003\%$ sample. The resulting energy, $\Delta
E_{sp-d}$, isolates the exchange splitting, plotted as a function of
$B$ in Fig. 4d along with a fit of the form $x N_0 (\alpha - \beta)
\langle S_x \rangle$.  While a fit giving $N_0 \beta=-3.5\pm1.5$ eV
closely approximates the data for $|B| < 5$ T, the description breaks
down at larger $B$.

Similar non-linear behavior in $\Delta E_{PL}$ at high fields in
samples with $d=$ 3, 5, and 7.5 nm gives a large uncertainty in our
estimates of $N_0 \beta$.  Further complicating the determination of
$N_0 \beta$ are the widely differing values extracted for samples of
different widths. Using fits similar to those shown in Fig.
\ref{fig4}c we find $N_0 \beta=+17.7\pm1.6$, $+98.0\pm7.3$, and
$-11.5\pm6.1$ eV for QWs with $d=$ 3, 5, and 7.5 nm, respectively.
Such dissagreement between samples indicates the incompleteness of our
model for the valance band; the mixing of valance band states may be
contributing to the problematic extraction of the $p-d$ exchange
coupling especially for small $d$ \cite{Snelling:1992}.  Clearly, more
work is necessary for the determination of $N_0 \beta$ in GaMnAs QWs
and its dependence on $d$.

In summary, strong evidence is presented of a direct relation between
the conduction band exchange constant and the electron kinetic energy
due to one-dimensional quantum confinement in GaMnAs QWs. It is a
quantitatively larger effect, but with the same sign, as what was
reported for II-VI DMS QWs \cite{Merkulov:1999}. Surprisingly, the
$s-d$ exchange coupling is antiferromagnetic in the QWs and
extrapolates to $-0.09\pm0.03$ eV in the limit of infinitely wide
wells indicating that antiferromagnetic $s-d$ exchange is a bulk
property of GaMnAs, a result which has not been predicted by current
DMS theories.

\begin{acknowledgments}
  This work was supported by DARPA, ONR, and NSF. One of us (N.P.S.)
  acknowledges support from the Fannie and John Hertz Foundation.
\end{acknowledgments}

\end{document}